\documentclass[twocolumn,twocolappendix]{aastex701}
\usepackage{hyperref,graphicx}
\usepackage{scalerel,stackengine}
\usepackage{enumitem}
\usepackage{amsmath,amssymb}

\begin{document}
\shorttitle{First Linearly Polarized Metric Solar Radio Emission}
\shortauthors{Dey et al.}

\title{First Robust Detection of Linear Polarization from Metric Solar Emissions: Challenging Established Paradigms}
\correspondingauthor{Soham Dey}
\email{sohamd943@gmail.com, sdey@ncra.tifr.res.in}

\author[0009-0006-3517-2031]{Soham Dey}
\affiliation{National Centre for Radio Astrophysics, Tata Institute of Fundamental Research, S. P. Pune University Campus, Pune 411007, India}
\email{sohamd943@gmail.com, sdey@ncra.tifr.res.in}
\author[0000-0001-8801-9635]{Devojyoti Kansabanik}
\affiliation{Cooperative Programs for the Advancement of Earth System Science, University Corporation for Atmospheric Research, 3090 Center Green Dr, Boulder, CO, USA 80301}
\email{dkansabanik@ucar.edu, devojyoti96@gmail.com}
\affiliation{The Johns Hopkins University Applied Physics Laboratory, 11001 Johns Hopkins Rd, Laurel, USA 20723}
\author[0000-0002-4768-9058]{Divya Oberoi}
\email{div@ncra.tifr.res.in}
\affiliation{National Centre for Radio Astrophysics, Tata Institute of Fundamental Research, S. P. Pune University Campus, Pune 411007, India}
\author[0000-0002-2325-5298]{Surajit Mondal}
\affiliation{Center for Solar-Terrestrial Research, New Jersey Institute of Technology, 323 M L King Jr Boulevard, Newark, NJ 07102-1982, USA}
\email{surajit.mondal@njit.edu}

\begin{abstract}
Polarimetric radio observations of the Sun can provide rich information about emission mechanisms and the propagation medium. For the past five decades, solar polarimetric studies at low radio frequencies have almost always assumed the absence of linear polarization. This has been based on the expectations from coronal propagation effects. Here we present the first robust evidence of linear polarization from solar emissions at meter wavelengths using simultaneous measurements with two telescopes of very different designs separated by thousands of kilometers -- the Murchison Widefield Array and the upgraded Giant Metrewave Radio Telescope. Both datasets show consistent linear polarization fractions, confirming this detection. Rapid changes in morphology, as well as the fractional linear polarization at small time and frequency spans, further rule out any possibilities of an instrumental origin. Assuming the absence of linear polarization in solar radio emissions can result in incorrect interpretation of solar observations as well as those of other flare stars, which are often guided by learnings from solar studies. This discovery highlights the need for relaxing this assumption and is essential for precise estimation of polarization signatures, ultimately leading to a better understanding of the plasma conditions in the Sun and other stars.
\end{abstract}


\section{Introduction}\label{sec:intro}
The solar corona, the outermost layer of the solar atmosphere, serves as a natural laboratory for investigating magnetized plasma under conditions difficult to achieve in terrestrial laboratories. Solar radio bursts — transient, intense emissions — are a direct manifestation of dynamic processes such as particle acceleration, plasma instabilities, and magnetic reconnection \citep{zlotnik1994,gary2023}. Traditionally, these bursts have been broadly classified based on their appearance in the time-frequency plane \citep{wild1950I,wild1950II,wild1950III}, referred to as dynamic spectrum (DS). The ones of particular interest here are referred to as type-I and type-III bursts in the literature. Type‐I noise storms appear as persistent, broadband enhancements in the background emission, interspersed with short, narrowband bursts, known as type-I bursts. They are thought to arise from energetic electrons confined within closed magnetic loops \citep{melrose1980typeI}. In contrast, type‐III bursts manifest as brief, rapidly drifting streaks in DS that trace semi‐relativistic electron beams escaping along open magnetic field lines \citep{melrose1970,zheleznyakov1970}. These bursts not only carry information about their underlying plasma emission mechanism \citep{dulk1985} but also bear the imprints of the electron density distribution and magnetic field configuration encountered during propagation. Polarimetric observations will enable us to study some properties of the emission mechanism and the medium better, such as polarities of magnetic fields, distribution of non-thermal electrons, and length scales of inhomogeneities, as compared to using total intensity (Stokes I) alone.

Propagation effects due to density irregularities and variable magnetic field orientations and strengths can alter the polarization state and even depolarize the signal. For example, noise storms and type-I bursts tend to be less circularly polarized near the solar limb than at the central meridian \citep{kai1962}, while type-III bursts, though theoretically expected to be highly circularly polarized, are often observed with weak ($<30\%$) or even negligible polarization \citep{dulk1980}. This reduced polarization is often attributed to mode coupling between magnetoionic modes in the corona, a process that may be further enhanced by large-amplitude Alfv\'en waves twisting the magnetic field lines \citep{melrose1974}. In such scenarios, radiation that is initially 100\% polarized in one magneto-ionic mode (ordinary or O-mode for plasma emission) can acquire a partial linear component upon traversing a quasi-transverse region, where the wave vector is nearly perpendicular to the magnetic field \citep{melrose1970,zheleznyakov1970}. A detailed understanding of these propagation effects could transform polarimetric studies of solar radio emission into a powerful diagnostic tool for probing coronal inhomogeneities and magnetic fields. Full-polarization, spectroscopic snapshot interferometric imaging studies at meter wavelengths hold great promise for this purpose. 

Although linear polarization from active solar regions at microwave frequencies ($\sim 5$ GHz) has been reported \citep{alissandrakis1994}, no robust detection has been made at meter wavelengths until now.
In this work, we report the first robust detection of linearly polarized radio emission from metric solar radio emissions, which challenges a decades-old paradigm. The Letter is organized as follows. Section \ref{sec:problem} provides a brief overview of this decades-old paradigm in solar physics and related observational challenges. Details of observations and data analysis are presented in Sections \ref{sec:obs} and \ref{sec:data_analysis}, respectively. Section \ref{sec:results} presents the key results from this study, followed by a discussion about the possible physical origins of the linearly polarized emission in Section \ref{sec:physical_explanations}. We conclude and discuss future directions in Section \ref{sec:conclusion}.

\section{Observational Challenges and The Decades-old Paradigm}\label{sec:problem}
Full polarization studies, including robust and unbiased detection of both linear and circular polarization from metric solar radio emissions, remain rare due to the limited availability of suitable instruments and significant calibration challenges. The scarcity of polarized calibrators at low radio frequencies makes it hard to follow the traditional approach for polarimetric calibration employed at higher frequencies. In addition, the wide field-of-view (FoV) design of most low-frequency arrays further complicates calibration \citep{Kansabanik_principle_AIRCARS}. Moreover, the high flux density of the Sun can contaminate calibrator observations, and the calibrator's true polarization properties are also altered by Faraday rotation in the Earth's magnetized ionosphere \citep{kansabanik2025polcal}. 

Consequently, many studies have relied on the decades-old, longstanding assumption, driven by the expectation of high Faraday rotation in the corona, that any observed linear polarization at low radio frequencies is entirely instrumental \citep{grognard1973, raja2013, kumari2017, mccauley2019, morosan2022, ramesh2023}. Faraday rotation is the rotation of the plane of linear polarization when an electromagnetic wave traverses through a birefringent medium. It depends on the electron density and the magnetic field strength along the line of sight and is proportional to the square of the observing wavelength ($\lambda^2$). The latter makes the effect of Faraday rotation more pronounced at meter wavelengths. The Faraday rotation is most naturally described by the rotation measure (RM), which is defined as Faraday rotation per $\lambda^2$. Using a well-accepted coronal density profile \citep{Newkirk1961} and magnetic field model \citep{Dulk1978}, a radial ray originating at 1 solar radius above the photosphere yields RM of the order of $10^5~ rad~m^{-2}$. Such high RM implies that linear polarization signatures in solar radio emissions at meter wavelengths are expected to be essentially erased.

Three primary mechanisms contribute to this depolarization: (i) depth depolarization, where emission from different coronal heights accumulates varying rotation angles, leading to a reduced vector sum; (ii) beam depolarization, caused by spatial fluctuations in RM across the telescope resolution, potentially canceling polarization vectors from adjacent sightlines; and (iii) bandwidth depolarization, arising from variations in RM-induced rotation across a finite frequency channel, effectively averaging out polarization within each spectral channel. Early claims from the late 1950s through the early 1970s reported linear polarization in type-III bursts \citep{cohen1959, akabane1961, kai1963, bhonsle1964, chin1971}. However, they were largely dismissed later as instrumental artifacts, due to inadequate calibration and the prevailing expectation of strong Faraday rotation in the corona \citep{grognard1973, boischot1975}. 

As a result, subsequent studies and even recent studies have either discarded the linearly polarized component or used it to devise calibration schemes designed to nullify it \citep{raja2013, kumari2017, mccauley2019, morosan2022, ramesh2023}. Here, we challenge this decades-old paradigm by reporting the first robust detection of linear polarization at meter wavelengths using spectro-polarimetric snapshots imaging observations from two distinct classes of solar radio bursts - type-I and type-III. 

\begin{figure*}[!htbp]
\centering
\includegraphics[trim={0.5cm 0.5cm 1cm 0.5cm},clip,width=0.5\linewidth]{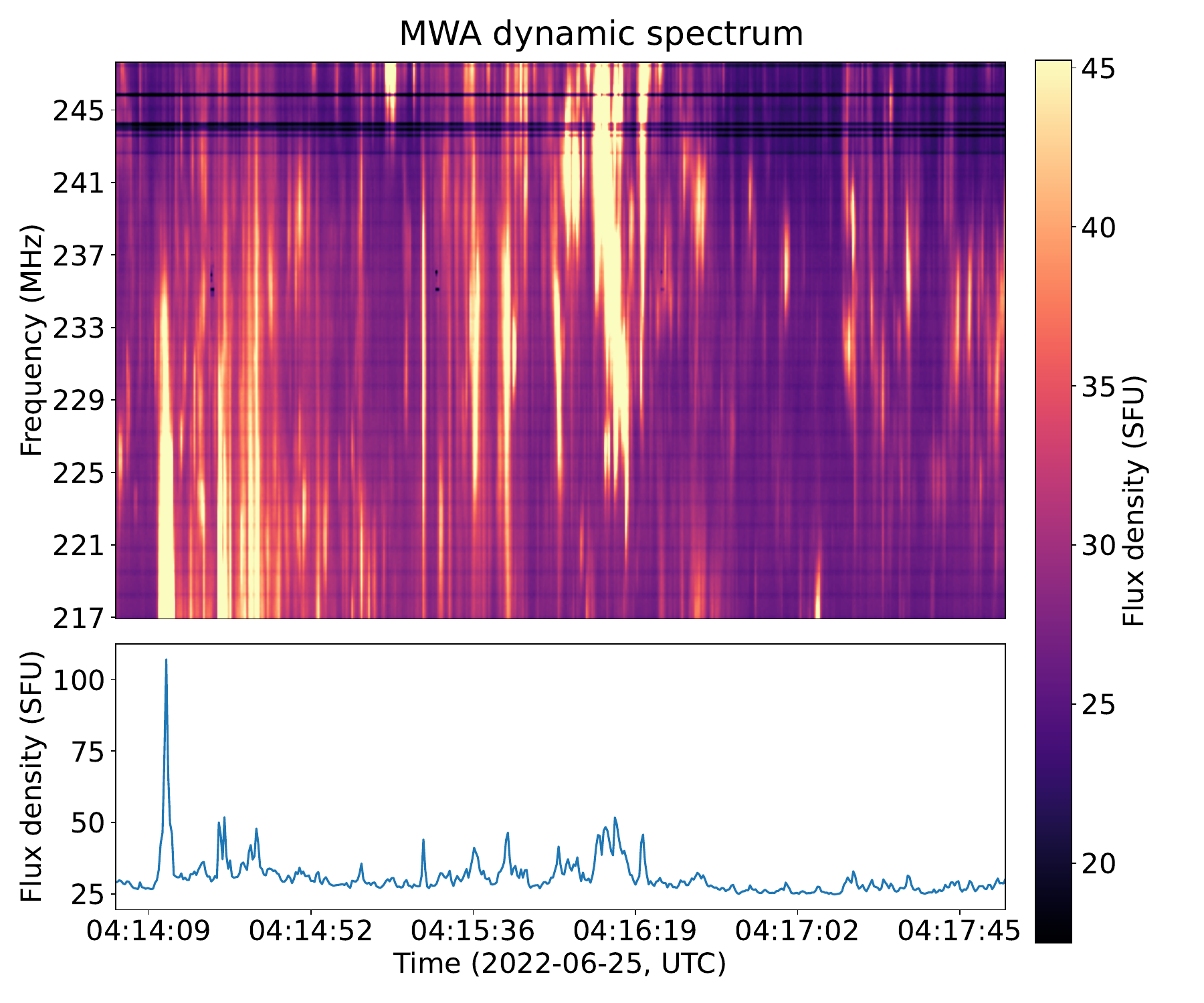}\includegraphics[trim={0.5cm 0.5cm 1cm 0.5cm},clip,width=0.5\linewidth]{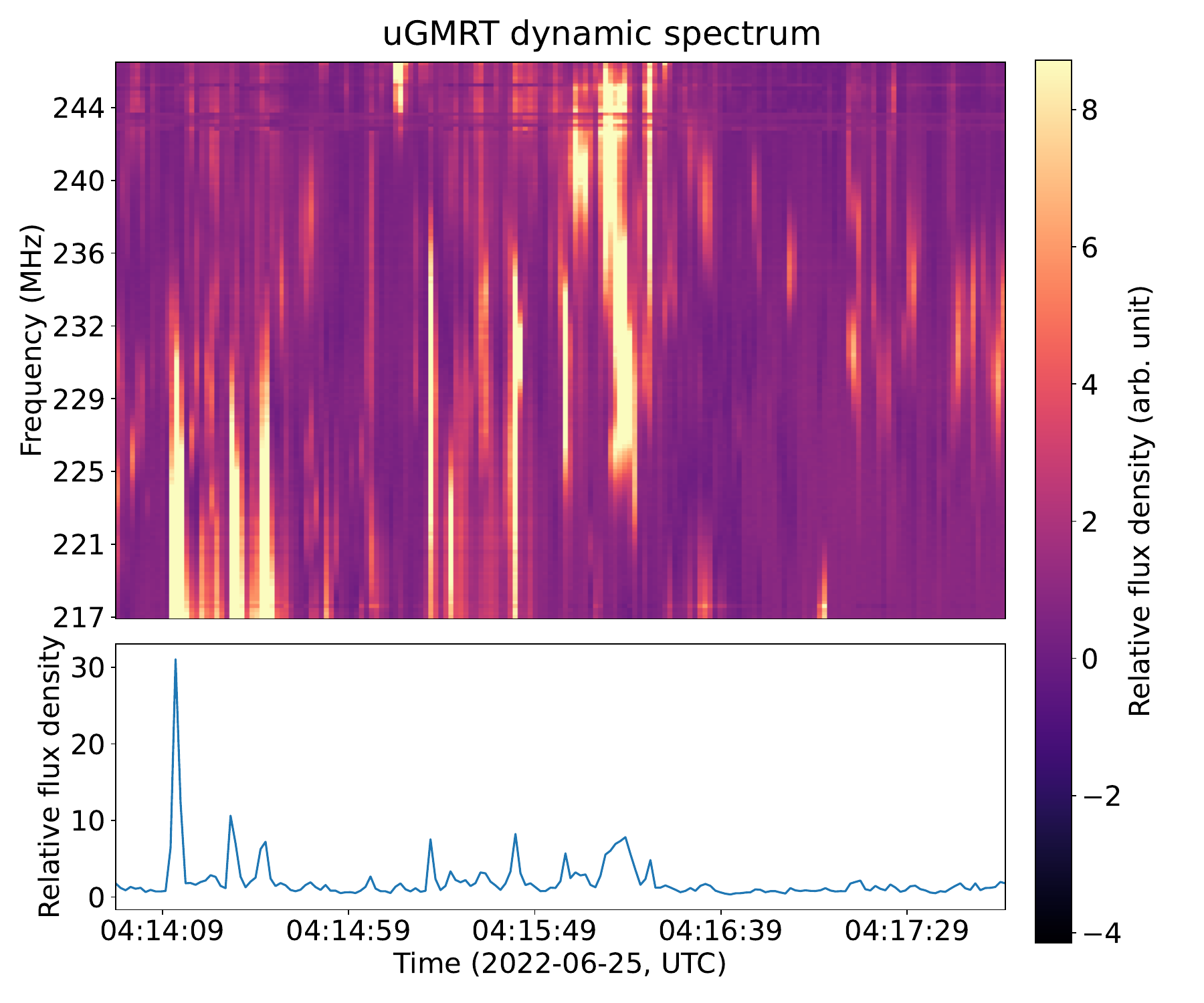}
\caption{The figure shows the Stokes I (total intensity) dynamic spectra (top panels) and the time series of the band-averaged flux density (bottom panels) for four minutes of the event on 25 June 2022, captured by the MWA (left panels) and uGMRT (right panels). A type-III burst took place at 04:14:14 UTC, which corresponds to the highest peak in the bottom panels.}
\label{fig:mwa_gmrt_ds}
\end{figure*}

\section{Observation Details}\label{sec:obs}
This study is based on simultaneous Murchison Widefield Array \citep[MWA;][]{Tingay2013} and upgraded Giant Metrewave Radio Telescope \citep[uGMRT;][]{Gupta2017} observations from 03:30 to 04:30 UTC on 2022 June 25, during which four NOAA active regions were present on the solar disk\footnote{\href{https://www.solarmonitor.org/?date=20220625}{NOAA active regions on 25 June 2022}}. Both instruments, designed for faint astronomical sources, require attenuation to observe the Sun. While both support full-Stokes observations, the MWA uses a linear polarization basis (X/Y) and uGMRT a circular basis (R/L), leading to different manifestations of instrumental polarization \citep{hamaker1996}. The MWA, located in Western Australia, operated in its phase-II configuration \citep{wayth2018} with 136 tiles over $\sim$5 km and a 30.72 MHz instantaneous bandwidth. It cycled across 129–247 MHz in 30.72 MHz chunks to match the uGMRT band-2 range, with each scan lasting 4 minutes. This study uses MWA data from 04:13:58–04:17:58 UTC over 217–247 MHz, capturing a type I noise storm and type III bursts (Figure~\ref{fig:mwa_gmrt_ds}). Observations were recorded by the MWAX correlator \citep{morrison2023} at 0.25 s and 10 kHz resolution, using a 10 dB attenuator. Calibrators (Centaurus-A, Hydra-A) were observed without attenuation outside solar hours.

The uGMRT, located near Pune, India, comprises thirty 45-m antennas spanning $\sim$25 km and offers arcsecond-level resolution at metre wavelengths. Observations were made using band-2 (120–250 MHz) with a 30 dB attenuator. Data were recorded using the GMRT Wideband Backend \citep{reddy2017} with 195.3 kHz frequency and 1.3 s time resolution. Full-Stokes visibilities (RR, RL, LR, LL) were obtained. The bright calibrator 3C48 (flux $\sim$42 Jy at band-2; \citealt{perley2017}) was used for bandpass, absolute flux, and instrumental polarization calibration. It was observed with the same attenuation, and its high flux ensured sufficient SNR. No phase calibrator was used to reduce overheads.

\section{Data Analysis}\label{sec:data_analysis}
\subsection{Calibration Procedure}\label{subsec:calibration}
The MWA and uGMRT observations were calibrated using different approaches and algorithms tailored for the respective instruments. Imaging of the data from the two instruments was also carried out independently. Data analysis for both instruments is performed primarily using Common Astronomy Software Applications \citep[CASA;][]{bean2022} and WSClean \citep{offringa2014}. The MWA calibration is done following the algorithms developed in \citet{kansabanik2025polcal,Kansabanik2022_paircarsI}. A detailed description is presented in Appendix \ref{sec:mwa_cal}. The uGMRT observations are calibrated using CASA, which is described in detail in Appendix \ref{sec:uGMRT_analysis}.

\subsection{Imaging Procedure}\label{subsec:imaging}
The dense {\it uv}-coverage of the MWA enables high time and frequency resolution spectroscopic snapshot imaging; however, to improve the signal-to-noise ratio of the weak linearly polarized emission, we averaged the data over 0.5 s and 160 kHz intervals. In contrast, due to the sparse {\it uv}-coverage of the uGMRT — particularly at short baselines ($<$150~$\lambda$), which introduce artifacts due to high-amplitude visibilities — we restricted imaging of the uGMRT dataset to baselines $>150\ \lambda$ and averaged the data over 10 s and 1.9 MHz. These choices balance improved {\it uv}-sampling and signal to noise with the need to preserve the intrinsic spectral and temporal variability of the solar emission. A manual mask based on Stokes~I dirty images was applied, and the same image weighting scheme as used for the MWA was adopted.

\subsection{Calculation of Linear Polarization Fraction and Error Estimation}\label{subsec:polfrac_cal}
To estimate the linear polarization fraction, only regions where Stokes I was greater than $10\sigma$, and then any regions above 5$\sigma$ in Stokes Q and U were considered, where $\sigma$ is the rms noise of the corresponding Stokes images. 
For each source region, we did as follows:
\begin{itemize}
    \item We searched the location of peak linear polarization signal within that region, given by the maximum of $\sqrt{Q^2+U^2}$.
    \item We considered a point spread function (PSF) sized region centered on the evaluated peak polarization signal.
    \item Within this PSF-sized region, the mean values of Stokes I, Q, and U were computed.
    \item This was then used to estimate the linear polarization fraction, defined as $f_L = \sqrt{Q^2+U^2}/I$
\end{itemize}
Although the rms noise of individual Stokes follows a Gaussian distribution, linear polarized intensity $L$ and its fraction $f_L$, do not. Hence, we used a Markov Chain Monte Carlo \citep[MCMC;][]{foremanmackey2013} based approach to estimate the errors on $L$ and $f_L$, which is described in detail in Appendix \ref{sec:error}. 

\section{Results}\label{sec:results}
\subsection{Detection of Linearly Polarized Emission}\label{subsec:detection}
The simultaneous observations presented in this work at overlapping frequencies with the MWA and the uGMRT, combined with a series of continuous developments of state-of-the-art calibration algorithms optimized for solar observations \citep{Mondal2019,Kansabanik_principle_AIRCARS,Kansabanik2022_paircarsI,Kansabanik_paircars_2,kansabanik2025polcal}, allow us to demonstrate that meter-wavelength solar radio emission can show a significant intrinsic linear polarization component.

\begin{figure*}[!htbp]
\includegraphics[width=\linewidth]{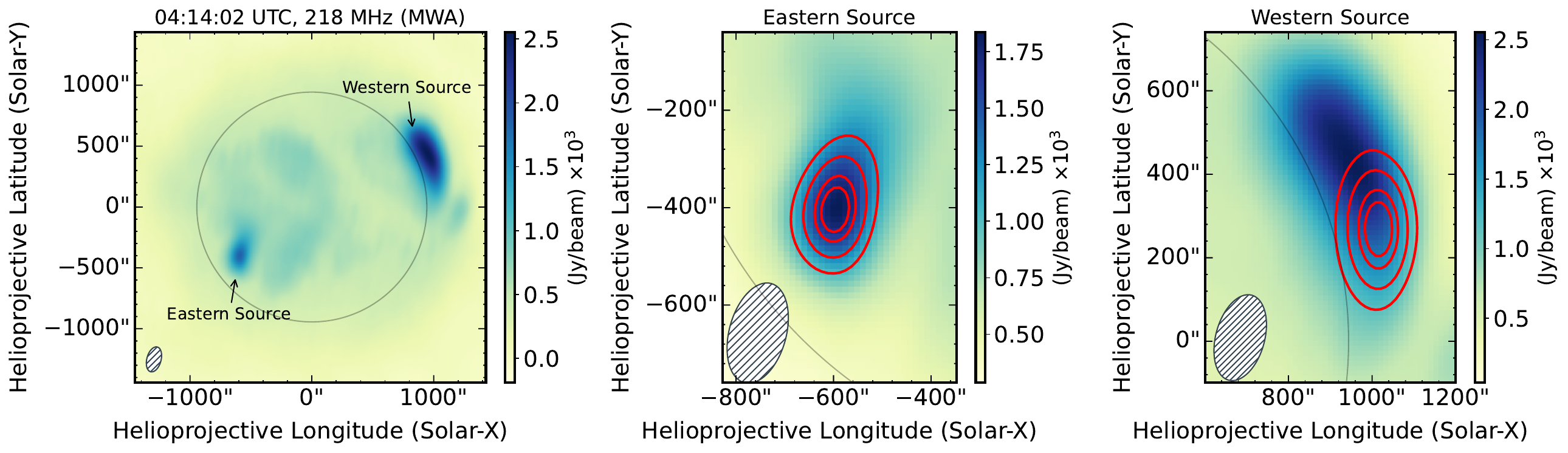}
\includegraphics[width=\linewidth]{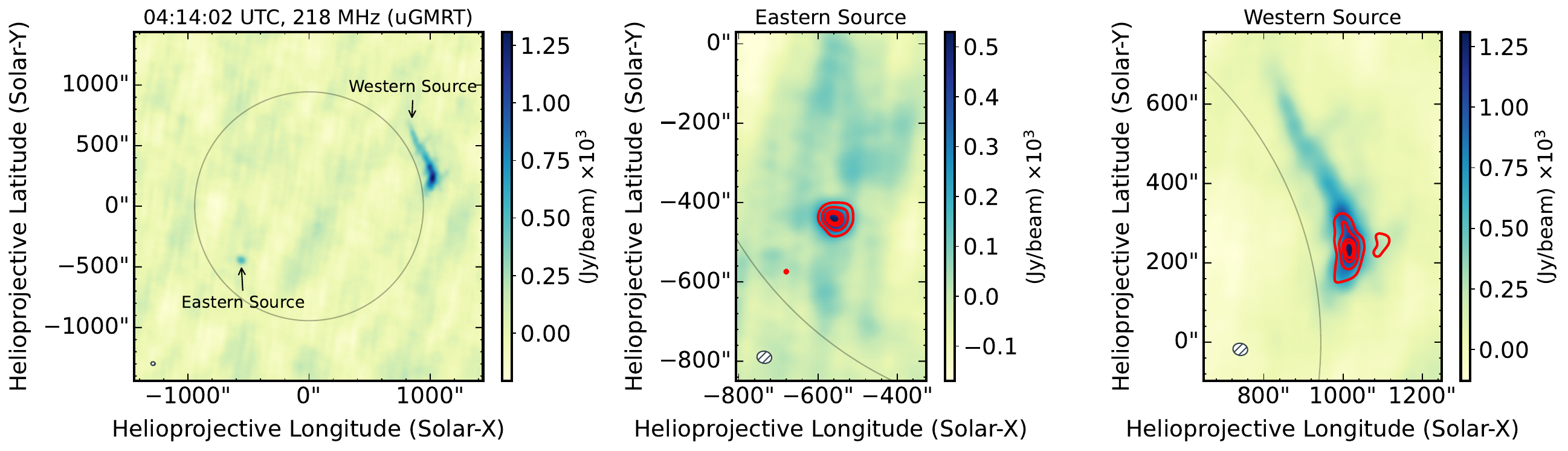}
\caption{Simultaneous detection of linearly polarized emission at 218\,MHz from the MWA (top panels) and uGMRT (bottom panels). Two bright radio sources on the eastern and western limbs are marked in the left panels, with zoomed-in views shown in the middle and right panels. Red contours denote linear polarization intensity ($L$) at 0.4, 0.6, 0.8, and 0.9 of the peak, and black circles indicate the optical solar disk. Both sources show partial linear polarization. The eastern source, weaker in Stokes~$I$, has a polarization fraction of $13.5 \pm 0.5\%$ (MWA) and $12.3 \pm 2.0\%$ (uGMRT). The brighter western source shows lower polarization: $6.0 \pm 0.3\%$ (MWA) and $5.9 \pm 0.5\%$ (uGMRT).} 
\label{fig:type_I_type_III_demon}
\end{figure*}

Figure \ref{fig:type_I_type_III_demon} illustrates a comparison of Stokes I intensity maps with linear polarization intensity ($L$) contours for a type-I noise storm at 218 MHz observed simultaneously by both instruments. 
The top panels represent the MWA observations, and the bottom panels represent the uGMRT observations. The MWA and uGMRT maps correspond to time integrations of 3 s and 10 s, respectively, and both these maps have been made over the 217-219 MHz band. In both images, linearly polarized sources are evident at the west (right) limb and the east (left) region of the Sun. Notably, the western source -- having a higher Stokes I intensity -- exhibits a lower linear polarization fraction ($f_L$) of $6.0 \pm 0.3\%$ with the MWA and $5.9 \pm 0.5\%$ with the uGMRT, whereas the eastern source shows an $f_L$ of $13.5 \pm 0.5\%$ with the MWA and $12.3 \pm 2.0\%$ with the uGMRT. 
The degree of circular polarization for these sources is substantially higher in comparison: $25.6 \pm 0.7\%$ for the western source and $48.9 \pm 1.3\%$ for the eastern source as estimated with the MWA.
\begin{figure*}[!htbp]
    \centering
    \includegraphics[width=0.93\linewidth]{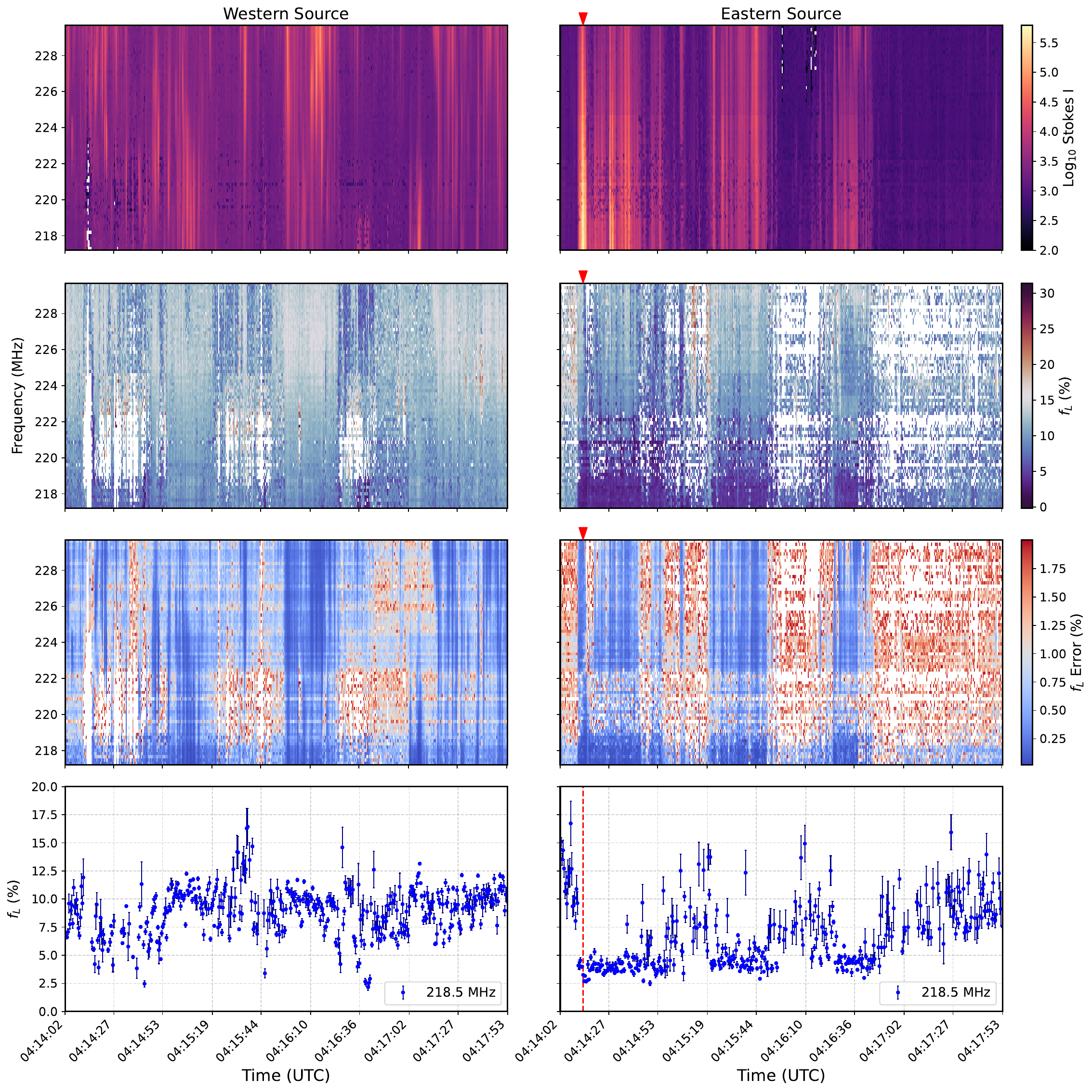}
    \caption{The figure shows the time-frequency variation of Stokes~$I$ and linear polarization fraction ($f_L$) for two noise storm sources. Rows~1--3 display Stokes~$I$, $f_L$, and its error, respectively, with white points indicating non-detections ($I < 10\sigma$ or $Q/U < 5\sigma$). The bottom row shows the time variation of $f_L$ at 218.5\,MHz. Rapid, source-specific changes in $f_L$ are evident. A type~III burst at 04:14:14\,UTC (marked by a red inverted triangle) in the eastern source leads to a sharp drop in $f_L$ from $>10\%$ to $<5\%$ at 217--220\,MHz. These distinct variations confirm that the observed polarization is intrinsic to the solar emission and not of instrumental origin.}
    \label{fig:dynspec_I_fl_efl}
\end{figure*}

\subsection{Validation of Observed Linear Polarization}\label{subsec:validation}
To confirm the validity of these unexpected findings, we present multiple independent lines of evidence -- (1) comparing simultaneous observations at overlapping frequencies from two very different telescopes; (2) examining the spectral and temporal variations in the linear polarization fraction ($f_L$) from different active emission sources on the solar disk; and (3) examining the evolution of polarization map during a type-III burst. These evidences, along with the arguments for its robustness, are presented next. 

The two telescopes used here are geographically separated by thousands of kilometers and have starkly different designs. The MWA is an aperture array instrument, located in Australia, with electronically steered elements distributed over a 5 km footprint and comprising groups of dipoles giving it a large FoV \citep{Tingay2013}. The uGMRT, located in India, comprises 30 large steerable parabolic dishes spread over distances of up to 25 km \citep{Gupta2017}. Calibration of these instruments requires different approaches primarily due to their different designs and ionospheric regimes. Hence, the observations were calibrated using different analysis strategies as detailed in Section \ref{subsec:calibration}. This implies that the images obtained from these instruments are highly unlikely to share similar systematics or artifacts. The consistent and independent detections from two very different and geographically well-separated telescopes provide compelling evidence that the observed linear polarization must be intrinsic to the solar radio emission incident on these telescopes.

Figure \ref{fig:dynspec_I_fl_efl} displays the spatially resolved DS, from the MWA observation. The top row shows the total intensity, the second row represents linear polarization fraction ($f_L)$, and the third row indicates the error in $f_L$. $f_L$ varies significantly over short time and frequency spans - from 2\% up to $\sim 31\%$ - with distinct spectro-temporal patterns for each source. In addition, for the eastern source, a type-III burst observed at 04:14:14 UTC lasting only one second is immediately accompanied by a sharp decline in linear polarization, from values above $10\%$ to consistently below $5\%$ in the 217-220 MHz range. Given the aperture array design of the MWA, characterized by its absence of moving parts and a simple design \citep{Tingay2013}, instrumental leakage is expected to vary smoothly and slowly across time and frequency. In addition, its wide  FoV ensures that instrumental primary beam leakage does not vary abruptly over the span of the solar disk \citep{Kansabanik2022_paircarsI}. Therefore, such localized rapid variations of polarization fraction seen in Figure \ref{fig:dynspec_I_fl_efl} cannot be attributed to residual instrumental polarization and must be intrinsic to the incident radiation.  

\begin{figure*}[!htbp]
\centering
 \includegraphics[width=0.75\linewidth, keepaspectratio]{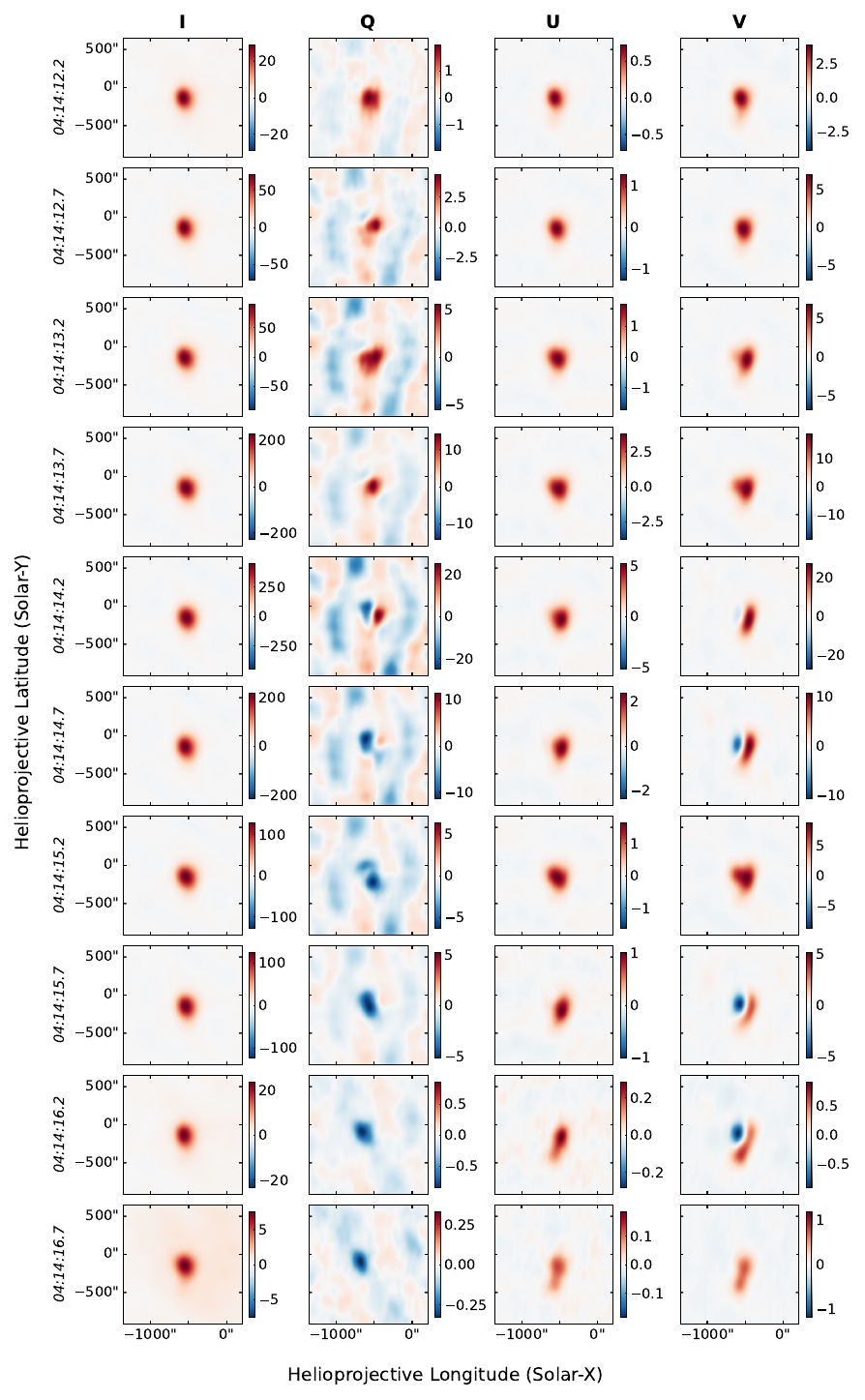}
\caption{Temporal evolution of polarization maps at 220 MHz for the eastern source during a type-III burst. The first column shows Stokes I (total intensity), followed by Stokes Q, U, and V. All of them are in units of kJy/beam. It is evident that at any given time slice these maps show uncorrelated morphologies across Stokes parameters. Notably, at 04:14:14.2 UTC, Stokes Q reverses sign, while Stokes U elongates — behaviors distinct from  Stokes I and V, confirming these features as being intrinsic to the solar emission rather than instrumental leakage.}
\label{fig:linpolvariation_type_III_tranposed}
\end{figure*}

Figure \ref{fig:linpolvariation_type_III_tranposed} illustrates the temporal evolution of polarization morphologies during the type-III burst for the eastern source region at 220 MHz. The first column displays the Stokes I (total intensity) map, followed by the linear polarization components Stokes Q and U, and the circular polarization component Stokes V in subsequent columns. Notably, the Stokes Q maps reveal striking dynamics: at 04:14:14.2 UTC, coinciding with the peak in Stokes I, the Stokes Q signal splits into distinct positive and negative lobes, with a clear transition from a predominantly positive signature before the burst to a negative one immediately after the burst. In contrast, the Stokes U maps always show only a positive feature that evolves into an elongated structure following the burst, while the Stokes V map also splits into positive and negative components and continues to evolve post-event. These complex and uncorrelated changes in morphologies across the polarization parameters are in stark contrast to the relatively small changes in structures observed in the Stokes I map. Such disparate behavior indicates that the observed patterns are intrinsic to the incident solar radiation rather than resulting from instrumental polarization (leakages from Stokes I to other Stokes parameters and from Stokes V to Stokes U), which can only produce similar morphological signatures across the Stokes parameters.

\section{Possible Physical Origins}\label{sec:physical_explanations}
Having ruled out an instrumental origin for the observed linear polarization, its unexpected presence warrants a deeper exploration of its origins. While the types of solar radio bursts reported here are known to exhibit circular polarization \citep{dulk1980,mugundhan2018,rahman2020,morosan2022}, the existence of a linear polarization component highlights a significant gap in our current understanding.  Linearly polarized emission can either originate close to the emission region or result from propagation effects far away en route to the observer. If generated close to the emission region, the prevailing conditions must permit only weak depolarization. Scattering and Faraday depolarization, previously thought to dominate at these wavelengths, may be less effective under certain conditions. A recent study of this same dataset suggests that the observed sizes of type-I sources are $\sim 3 - 5$ times smaller than predictions from widely accepted coronal scattering models \citep{mondal2024}. This discrepancy indicates that the coronal environment is far more complex than often assumed, underscoring the limitations of existing scattering and coronal models.

In the complex magnetic environments near active regions where these emissions occur, the perpendicular magnetic field component may play a significant role and impact its polarization. The detailed evolution of the topology of the magnetic field and its strength in the coronal active regions remains hard to constrain. The most intriguing scenario arises when the wave vector is perpendicular to the magnetic field (quasi-transverse propagation). Such situations can arise in multiple ways - the magnetic field itself changes direction \citep{bastian1995}, small-scale density inhomogeneities alter the local refractive index, changing the direction of the wave vector \citep{bastian1995}, and the presence of current sheets \citep{gopalswamy1994}. In such scenarios, mode coupling between the magneto-ionic waves occurs and can give rise to linear polarization. The discrepancy between the low observed circular polarization and the theoretically expected high values of type-III bursts is often attributed to these phenomena \citep{melrose1974}. 

Another intriguing possibility involves the generation of linear polarization through reflections from interfaces with significant density contrast. To explain the observations of linear polarization observed from the flare star UV Ceti, such a scenario was suggested involving over-dense plasma regions at much higher coronal heights in order to significantly reduce the effect of coronal Faraday depolarization \citep{bastian2022}. If the angle of incidence approaches Brewster’s angle, significant linear polarization can result. Such density structures are known to exist in the solar corona \citep{Poirier2020}, suggesting that such a mechanism can be operational here.

During the type-III burst, the polarization maps exhibit rapid, complex variations within seconds, with a positive structure first becoming bipolar and then turning negative (Figure \ref{fig:linpolvariation_type_III_tranposed}). Such dynamic changes in the morphology of the linearly polarized emission are more likely to arise due to localized changes in coronal electron density and pitch angle of the electron beam. Furthermore, the distinctly different morphologies observed in Stokes Q and U suggest the presence of highly intricate magnetic field structures in the emission and/or propagation regions. Thus, it seems that the scenario proposed for UV Ceti, reflection from interfaces of significant density contrast, is unlikely to be the primary mechanism of generation of linear polarization in this case.

\section{Conclusion}\label{sec:conclusion}
Regardless of the precise mechanism, the persistence of linear polarization points to significant gaps in our current understanding of the solar corona, coronal radio emission, and wave propagation through its magnetized plasma. Future studies must account for the presence of linear polarization, rather than dismissing it based on an assumption which, as we have shown, is not assured to always hold. 
Ignoring the presence of Stokes Q and U, or relying on calibration approaches that set them to zero, may lead to inaccurate polarization measurements.
In addition, measurements of Stokes Q and U not only open up an unexplored part of the phase space, but also provide additional independent measurables to help constrain the physics of the system. 

Insights drawn from this work are relevant beyond the Sun to other astrophysical contexts, such as flare stars, where direct analogies with solar radio bursts are often drawn to infer the underlying emission mechanisms \citep{melrose1993, lynch2017,callingham2021}. Stellar radio bursts often exhibit coherent emission processes similar to those as solar radio bursts. Yet plasma emission \citep{dulk1985} is frequently dismissed in favor of electron cyclotron maser emission \citep{treumann2006}, with detection of linear polarized component \citep{lynch2017,callingham2021} cited as one of the key evidences as such emission is deemed to be absent in the solar case.

This paradigm-shifting discovery opens new avenues for investigating wave propagation in magnetized coronal plasmas, yielding an informative new observable that has been ignored for decades. By unraveling the origins and implications of linear polarization in solar radio bursts, we will refine our understanding of the solar corona and its magnetic field and extend this knowledge to broader astrophysical phenomena.

\begin{acknowledgements}
This scientific work uses data obtained from Inyarrimanha Ilgari Bundara, the CSIRO Murchison Radio-astronomy Observatory. We acknowledge the Wajarri Yamaji People as the Traditional Owners and Native Title Holders of the observatory site. Support for the operation of the MWA is provided by the Australian Government (NCRIS), under a contract to Curtin University administered by Astronomy Australia Limited. We acknowledge the Pawsey Supercomputing Centre which is supported by the Western Australian and Australian Governments. We thank the staff of the GMRT who have made these observations possible. The GMRT is run by the National Centre for Radio Astrophysics of the Tata Institute of Fundamental Research. S.D. and D.O. acknowledge support from the Department of Atomic Energy, under project 12-R\&D-TFR-5.02-0700.
\end{acknowledgements}

\appendix
\restartappendixnumbering 
We discuss details of the calibration procedure employed for both the MWA and uGMRT observations in these appendices.\\

\section{Calibration of MWA Observation}\label{sec:mwa_cal}
The MWA dataset was calibrated using a Centaurus-A observation taken before sunrise. While the native spectral resolution of the dataset is 10 kHz, it was averaged to 160 kHz for further analysis. Standard flagging was first performed. Then, for each 160 kHz spectral channel, we estimated complex gains and cross-hand phases, where the cross-hand phase is the phase difference between signals from two orthogonal polarizations. These calibration solutions were then applied to the solar data. Next, self-calibration was performed independently on each 160 kHz channel of the solar data and then applied to get the spectroscopic snapshot images at 0.5 s and 160 kHz integrations. Subsequently, the necessary primary beam corrections and the leakage from Stokes I to the other Stokes parameters were estimated, which is known as {\it polconversion} \citep{Hamaker2000}. These were then applied to the images to get the final corrected images. The procedure adopted for cross-phase calibration and polconversion correction is discussed next.

\subsection{Cross-hand Phase Calibration of the MWA}\label{subsec:mwa_crossphasecalib}
MWA observes in the linear polarization basis (X and Y). Hence, improper estimation of the cross-hand phase can lead to the mixing of Stokes U and V. Multiple instances of highly circularly polarized active solar emission have been reported \citep{dulk1980,benz1985}. Hence, precise cross-hand phase calibration is necessary to ensure that no spurious Stokes U sources arise due to the presence of strong Stokes V features. We have used the formalism developed by \cite{kansabanik2025polcal} for low-frequency polarization calibration using unpolarized sky.

\begin{figure*}
    \centering
    \includegraphics[width=\linewidth]{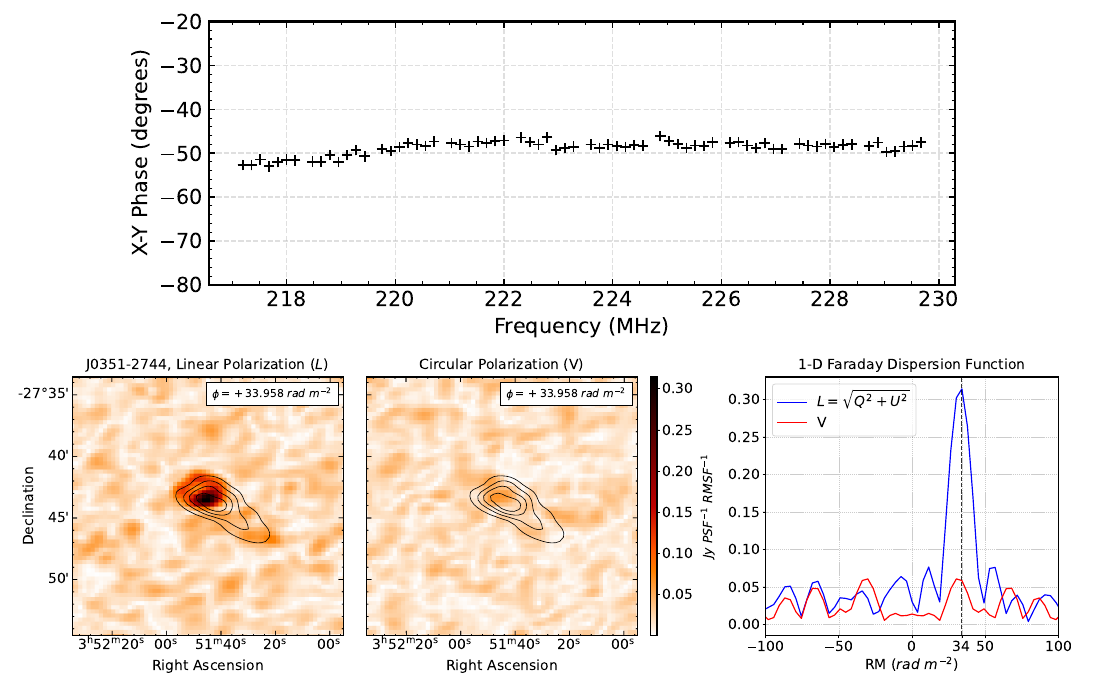}
    \caption{The top panel shows the cross-hand phase variation with frequency from the Centaurus A field. The bottom panels validate these solutions using J0351-2744. Linear and circular polarization maps at +33.958 rad m$^{-2}$ are shown (bottom left and middle), with Stokes I contours at 20\%, 40\%, 60\%, and 80\% of the peak. The bottom right shows the 1-D Faraday dispersion function for the eastern hotspot, peaking at +33.958 rad m$^{-2}$ (blue). Residual circular polarization is at the noise level. The small offset from the catalog RM (+33.58 rad m$^{-2}$) is likely due to uncorrected ionospheric RM.}
    \label{fig:crossphase_verification}
\end{figure*}

Briefly, we used the sky model of the Centaurus A field, which was also employed for estimating the initial complex gains. Centaurus A is known to be strongly depolarized at lower frequencies, resulting in an effectively zero fractional polarization at our observing frequency \citep{osullivan2013}. The MWA primary beam, which is highly polarized, leads to the apparent observed visibilities of Centaurus A being polarized. We applied the full embedded element MWA tile beam model \citep{sokolowski2017} to estimate the polarized apparent model visibilities. Using this apparent polarized model, the bandpass is estimated. Next, the cross-hand phase was estimated by minimizing a minimization function obtained from the bandpass corrected and model visibilities. The estimated cross-hand phases for each channel are shown in the top panel of Figure \ref{fig:crossphase_verification}.

To validate the cross-hand phase calibration, we have followed a similar approach taken in \cite{kansabanik2025polcal}. We applied the estimated cross-hand phase values per spectral channel to observations of the active galactic nucleus (AGN) GLEAM J035140-274354 from the same day. AGNs are known to exhibit significant linear polarization.  Hence, they provide a suitable test source for assessing the accuracy of our calibration procedure. For this validation exercise, we focus on the eastern hotspot of J0351-2744, which is known to have a Rotation Measure (RM) of +33.58 rad/m$^2$ \citep{bernardi2013}. This RM is sufficient to cause the Stokes U polarization intensity of the source to vary across the frequency band of observation. If the cross-hand phase calibration is wrong, it will lead to a spectrally varying leakage into Stokes V with a peak at a Faraday depth of $\sim$ +33.58 rad/m$^2$ in the Faraday Dispersion Function. 

We have performed RM synthesis analysis for J0351-2744 after applying the cross-phase solutions, which is shown in the bottom panels of Figure \ref{fig:crossphase_verification}. We see that the linear polarization intensity peaks at +33.9 rad/m$^2$. The observed difference from the catalog RM arises due to ionospheric RM \citep{oberoi2012}, which has not been subtracted. The residual Stokes V intensity at $\sim$+33.9 rad/m$^2$ is comparable to the noise across Faraday depth. This verifies that the cross-hand phase calibration determined here is correct, and we applied the same to the solar data. 

\subsection{Image-based Pol-conversion Correction}\label{imagebasedleakcorr}
The primary beam corrections were done for each coarse channel of 1.28 MHz using the Full-Embedded Element beam model \citep{sokolowski2017}, which corrects for most of the instrumental polarization leakage. However, deviations of true beam response from the ideal beam require us to correct for the residual leakages. The image-based leakage was calculated on the self-calibrated and beam model corrected images for each 160 kHz, following the principles employed in the MWA polarimetric observations of astrophysical objects \citep{lenc2017} as well as the Sun  \citep{Kansabanik2022_paircarsI,Kansabanik_paircars_2}. 

\begin{figure*}[!htbp]
    \centering
    \includegraphics[width=0.8\linewidth]{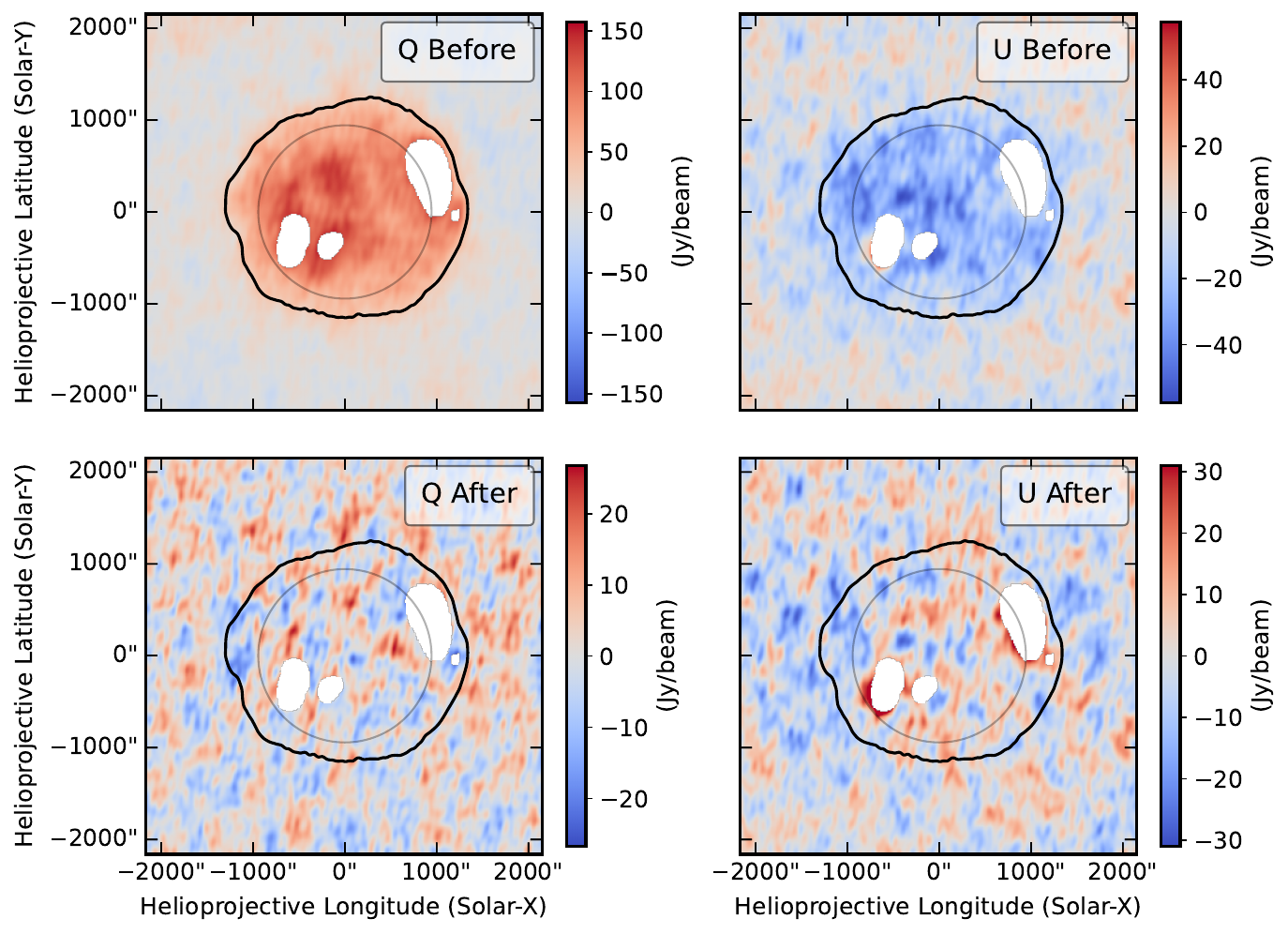}
    \caption{Comparison of Stokes I to Q, and I to U leakage before and after the image-based leakage correction. The top panel shows the Stokes Q (left) and U (right) maps at 217 MHz before image-based leakage corrections were done. Regions with brightness temperature $>10^6$ K have been masked. The bottom panels show the corresponding Stokes Q and U maps after the application of image-based leakage correction. The black contour on each of the maps indicates the 5-$\sigma$ level in Stokes I. After applying this correction, the mean values of Q and U in the quiet Sun region are 1.82 Jy/beam and -0.11 Jy/beam, respectively, which are significantly lower than the rms values of 4.70 Jy/beam and 4.46 Jy/beam.}
    \label{fig:residual_leakage}
\end{figure*}

The residual leakages from Stokes I into Q and U after primary beam correction are illustrated in the top panels of Figure \ref{fig:residual_leakage}, and the bottom panels show the residual leakages after the image-based corrections. In these figures, active emissions are masked and not included in the estimation of the leakage fraction. As evident from these bottom panels, the mean values of the Stokes Q and U in the quiet Sun region are $1.82$ Jy/beam and $-0.11$ Jy/beam, respectively, which are significantly lower than the rms values of 4.70 Jy/beam and 4.46 Jy/beam. Regions of linear polarization were defined as those with intensity above $5\sigma$ in both Stokes Q and U images, but no such regions are present in the residual leakage map, excluding the active emissions. 

We define the residual leakage limit based on 3$\sigma$ limit as,
\begin{equation}
    L_\mathrm{res} <\left|\frac{3\times\sigma_\mathrm{Q,U}}{I_\mathrm{max}}\right|
    \label{eq:leakage_limit}
\end{equation}, 
where $\sigma_\mathrm{Q}$ and $\sigma_\mathrm{U}$ are the rms noise values of Stokes Q and U in Jy/beam and close to the Sun, respectively. $I_\mathrm{max}$ is the maximum pixel value in Jy/beam in the quiet-Sun regions. Following this, the estimated residual leakage limit after image-based correction for our observation is 2.1\% for Stokes Q and 0.6\% for Stokes U. 

\begin{figure*}
\centering
\includegraphics[width=0.9\linewidth]{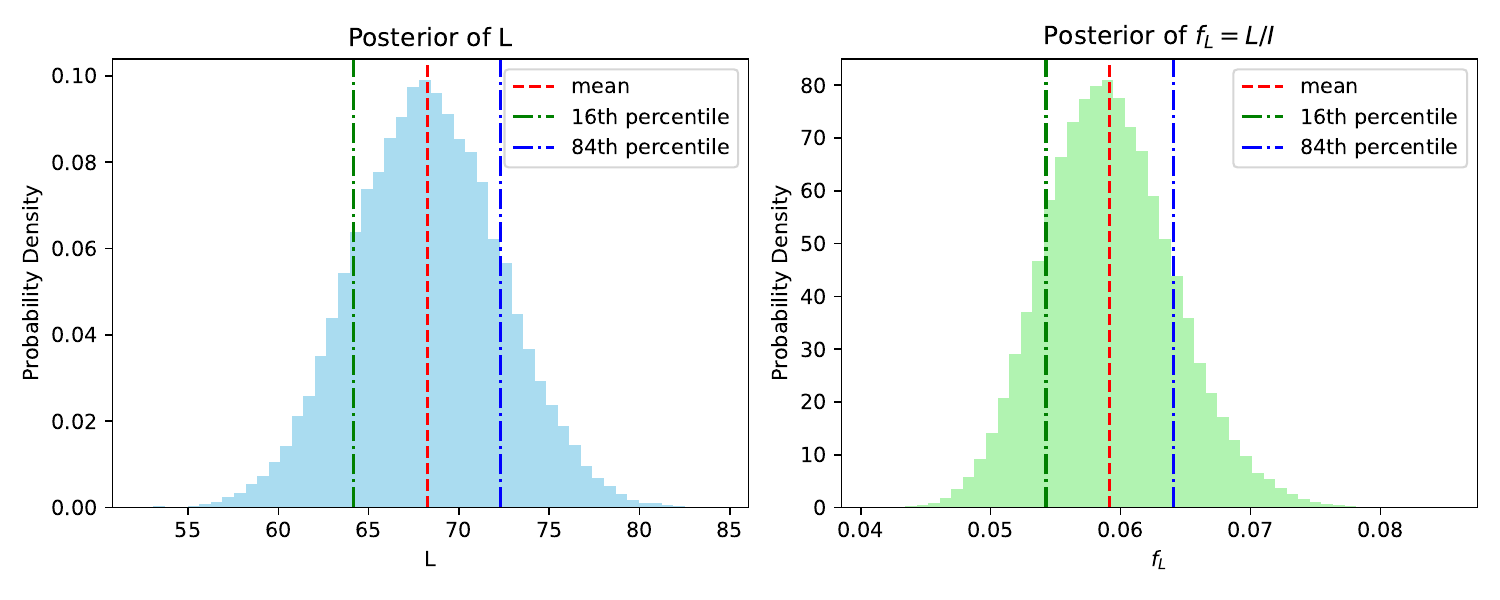}
\caption{The left and right panels show the posterior distributions of $L$ and $f_L$, respectively, of the western noise storm source from uGMRT observation at 218 MHz. The red vertical dashed line represents the mean, while the green and blue dash-dot lines represent the 16th and 84th percentiles, respectively. The uncertainty estimated throughout the paper is taken as half the difference between the 84th and 16th percentiles, which is 0.005 for $f_L$ for this specific instance.}
\label{fig:error_estimation}
\end{figure*}

\section{Calibration of uGMRT Observations}\label{sec:uGMRT_analysis}
Analysis of the uGMRT data was performed using CASA \citep{bean2022}. First, bad antennas and the radio frequency interference (RFI) affected spectral channels on both 3C48 and solar scans are removed. Next, we performed automated RFI-flagging on the uncalibrated 3C48 scan using the \textsf{tfcrop} algorithm available in \textsf{flagdata} task. Due to the RFI signals becoming increasingly decorrelated with increasing baseline lengths, the nature of observed RFI is different for short baselines and long baselines. Hence, we performed automated flagging on the baselines shorter and longer than 1 km, separately with different thresholds. The spectro-temporal variations in solar emission can mimic RFI in the time-frequency plane and make it hard for an automated RFI-flagging algorithm to distinguish between the two. Hence, we did not perform automated RFI-flagging on solar scans before calibration.

After the first round of flagging, the frequency-dependent complex gain of the instrument (instrumental bandshape) is estimated using the source model for 3C48 \citep{perley2017} and \textsf{bandpass} task. After applying the bandshape solutions on 3C48, we performed another round of automated flagging on the residual data using the \textsf{rflag} algorithm to flag low-level RFI. This was followed by some manual inspection to identify any residual low-level RFI and flag them as well from the four cross-correlation products. As the presence of residual RFI affects the determination of the instrumental band shape, it was re-determined after this round of flagging. 

Next, we used the \textsf{polcal} task to solve for frequency-dependent instrumental leakages, using 3C48 as an unpolarized calibrator. CASA follows a linearized formalism for estimating instrumental polarization \citep{hales2017}. This approximation is valid only when the instrumental leakages are small. For each spectral channel, we have flagged antennas that have an estimated instrumental polarization of more than 15\%. After removing these, we re-estimated instrumental polarization leakage and applied that to correct for leakage from Stokes I to other Stokes parameters.

We applied instrumental bandpass and frequency-dependent leakage estimates to the solar scans and then performed flagging on corrected data by manual inspection. We examined the visibility plane (referred to as the {\it uv}-plane in radio interferometry) to look for RFI-affected data for solar scans. Since the sky brightness distribution is smooth, the visibility distribution in the {\it uv} plane is expected to be smooth as well. For calibrated data, this makes it easier to identify outliers in the {\it uv}-plane for solar observations rather than searching for them in the time-frequency plane. These outliers were manually identified and flagged. Additionally, time or frequency slices with more than 80\% of their data flagged were flagged completely.

\section{Error Estimation of Linear Polarization Fraction}\label{sec:error}
The uncertainty in the $f_L$ is estimated by accounting for both the noise in the measurements and the statistical spread in the MCMC-derived value of $L$. First, the rms noise levels were determined from an off-source region in the Stokes I, Q, and U images, yielding $\sigma_\mathrm{I}, \sigma_\mathrm{Q}, \sigma_\mathrm{U}$. Q and U are modeled as $Lcos\phi$ and $Lsin\phi$, respectively, and a MCMC routine produced a posterior distribution for $L$ (shown in the left panel of Figure \ref{fig:error_estimation}) from which its uncertainty is taken as half the difference between the 84th and 16th percentiles. For each corresponding MCMC sample of $L$, the polarization fraction is computed as $f_L = L/I$. The spread in the posterior of $f_L$ values (shown in the right panels of Figure \ref{fig:error_estimation}), again determined from the 84th and 16th percentiles, gives the final error estimate in $f_L$.

\bibliography{ref}{}
\bibliographystyle{aasjournalv7}

\end{document}